\documentclass[aps,prl,twocolumn,preprintnumbers,superscriptaddress,amsmath,amssymb]{revtex4-1}
\usepackage[normalem]{ulem}
\usepackage{graphicx}
\usepackage{subfigure}
\usepackage{epsfig}
\usepackage{xcolor}
\usepackage{dcolumn}
\usepackage{bm}
\usepackage{ulem}
\usepackage{color}
\usepackage{amsmath}
\usepackage{amsfonts}
\usepackage{amssymb}

\usepackage{epstopdf}

\begin{document}

\title{Evidence for quasi-one-dimensional charge density wave in CuTe by angle-resolved photoemission spectroscopy}

\author{Kenan Zhang}
\affiliation{State Key Laboratory of Low Dimensional Quantum Physics and Department of Physics, Tsinghua University, Beijing 100084, China}

\author{Xiaoyu Liu}
\affiliation{State Key Laboratory of Low Dimensional Quantum Physics and Department of Physics, Tsinghua University, Beijing 100084, China}

\author{Haoxiong Zhang}
\affiliation{State Key Laboratory of Low Dimensional Quantum Physics and Department of Physics, Tsinghua University, Beijing 100084, China}

\author{Ke Deng}
\affiliation{State Key Laboratory of Low Dimensional Quantum Physics and Department of Physics, Tsinghua University, Beijing 100084, China}

\author{Mingzhe Yan}
\affiliation{State Key Laboratory of Low Dimensional Quantum Physics and Department of Physics, Tsinghua University, Beijing 100084, China}

\author{Wei Yao}
\affiliation{State Key Laboratory of Low Dimensional Quantum Physics and Department of Physics, Tsinghua University, Beijing 100084, China}

\author{Mingtian Zheng}
\affiliation{Hiroshima Synchrotron Radiation Center, Hiroshima University, Higashihiroshima, Hiroshima 739-0046, Japan}

\author{Eike F. Schwier}
\affiliation{Hiroshima Synchrotron Radiation Center, Hiroshima University, Higashihiroshima, Hiroshima 739-0046, Japan}

\author{Kenya Shimada}
\affiliation{Hiroshima Synchrotron Radiation Center, Hiroshima University, Higashihiroshima, Hiroshima 739-0046, Japan}

\author{Jonathan D. Denlinger}
\affiliation{Advanced Light Source, Lawrence Berkeley National Laboratory, Berkeley, California 94720, USA}

\author{Yang Wu}
\affiliation{Department of Physics and Tsinghua-Foxconn Nanotechnology Research Center, Tsinghua University, Beijing, 100084, China}

\author{Wenhui Duan}
\affiliation{State Key Laboratory of Low Dimensional Quantum Physics and Department of Physics, Tsinghua University, Beijing 100084, China}
\affiliation{Collaborative Innovation Center of Quantum Matter, Beijing, China}

\author{Shuyun Zhou}
\altaffiliation{Correspondence should be sent to syzhou@mail.tsinghua.edu.cn}
\affiliation{State Key Laboratory of Low Dimensional Quantum Physics and Department of Physics, Tsinghua University, Beijing 100084, China}
\affiliation{Collaborative Innovation Center of Quantum Matter, Beijing, China}

\begin{abstract}

{\bf We report the electronic structure of CuTe with a high charge density wave (CDW) transition temperature T$_c$ = 335 K by angle-resolved photoemission spectroscopy (ARPES). An anisotropic charge density wave gap with a maximum value of 190 meV is observed in the quasi-one-dimensional band formed by Te $p_x$ orbitals. The CDW gap can be filled by increasing temperature or electron doping through {\it in situ} potassium deposition. Combining the experimental results with calculated electron scattering susceptibility and phonon dispersion, we suggest that both Fermi surface nesting and electron-phonon coupling play important roles in the emergence of the CDW.}
\end{abstract}

\maketitle

Low dimensional materials have the tendency to form new orderings with novel physics. Charge density wave (CDW) \cite{CDWrev} is one of the most fundamental phenomena and has been discovered in various transition metal dichalcogenides (TMDCs) \cite{Smith1985Band, Moncton1975Study} (e.g. TaSe$_2$ \cite{BergerTaSe2}, TaS$_2$ \cite{FengDLTaS2} and NbSe$_2$ \cite{NbSe2}) and rare-earth tritellurides RTe$_3$ \cite{Dimasi1995Chemical, Sacchetti2007Pressure}(R = rare earth elements, e.g. CeTe$_3$ \cite{Brouet2004Fermi} and SmTe$_3$ \cite{Gweon1998Direct}). In the CDW state \cite{TMDCDWrev, Gr2009Density}, an instability of the metallic Fermi surface involving electron-phonon interaction \cite{Rossnagel2005Fermi} or electron-electron scattering \cite{Straub1999Charge, Gweon1998Direct} leads to a modulation of the lattice coupled to changes in the conduction electron density in the real space with a period $\lambda_c$. Such modulation induces an energy gap at the Fermi wave vector $k_F$=$\pi / \lambda_c$, thereby lowering the electronic energy of the occupied electronic states in the CDW phase.

Copper-based chalcogenides form a large family with diverse properties and potential applications in solid-state devices such as thermoelectrics, batteries, and photovoltaics \cite{Li2000ChemInform,Chen1985Polycrystalline,Okimura1980Electrical}. Among these materials, the stoichiometric compound  CuTe exhibits a modulation of Te atoms, forming dimers and trimers at low temperature, suggesting a CDW transition \cite{Stolze2013CuTe}. Different from the quasi-2D TMDCs and RTe$_3$, Te atoms in CuTe  form quasi-1D chains, and therefore, CuTe can be viewed as a quasi-1D CDW system with Peierls-like distortion \cite{Peierls1955Quantum}.  So far, the electronic structure of CuTe, in particular the electronic structure signature of the CDW (e.g.  CDW gap, nesting vector etc.) as well as the underlying mechanism for CDW, has remained missing.

Here we provide direct experimental evidence for the CDW in CuTe by angle-resolved photoemission spectroscopy (ARPES) measurements.  An anisotropic CDW gap with a maximum value of 190 meV is observed in the quasi-1D band formed by Te $p_x$ orbitals through a nesting wave vector $q_x$ = 0.4 a$^*$ which matches well with previous diffraction results \cite{Stolze2013CuTe}.  By increasing the temperature or electron doping through {\it in situ} potassium deposition, the CDW gap is gradually filled and eventually disappears. These experimental results, combined with analysis from the calculated electron scattering susceptibility and phonon spectrum, suggest that both electron-phonon coupling and electron-electron interaction play important roles in the CDW formation in CuTe.

\begin{figure*}
\centering
\includegraphics[width=17.8 cm] {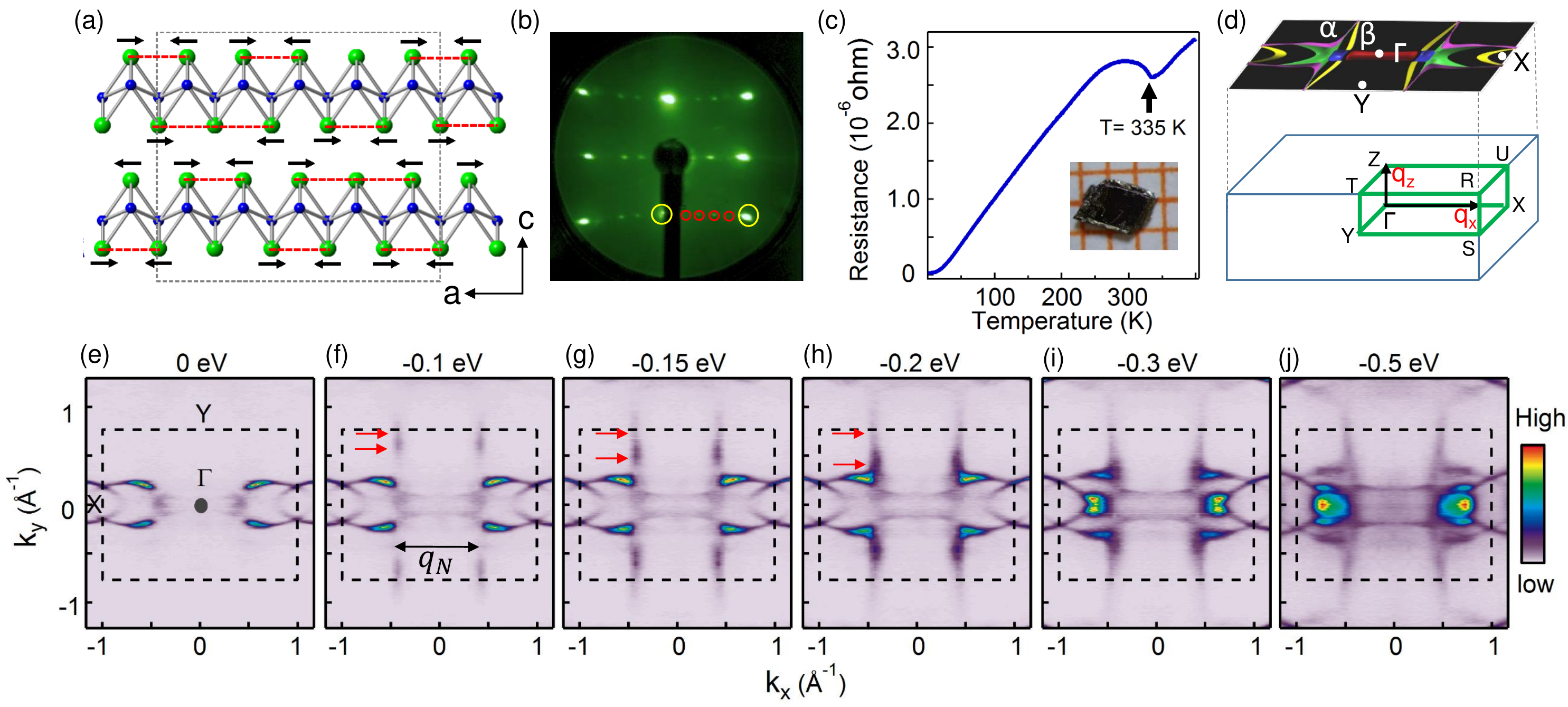}
\label{Fig1}
\caption{\textbf{(a)} Crystal structure of CuTe. Arrows indicate the movement of Te atoms in the CDW phase. \textbf{(b)} LEED pattern at 80 K shows Bragg peaks (yellow circle) and reconstruction peaks (red circle). \textbf{(c)} Temperature dependent resistivity of CuTe. The inset shows a picture of a typical single crystal of a few mm size. \textbf{(d)} Bulk Brillouin zone and calculated Fermi surface for the normal phase of CuTe with high-symmetry points labeled. \textbf{(e-j)} Intensity maps at constant energies from E$_F$ to -0.5 eV measured at 20 K, with photon energy of 80 eV and p-polarization. The black arrow indicates the nesting wave vector, and the red arrows mark the edges of the ungapped area.}\label{Fig1}
\end{figure*}

High quality CuTe single crystals were grown by self-flux method. Surface-sensitive ARPES measurements have been performed at BL 4.0.3 of the Advanced Light Source and  BL 1 \cite{Iwasawa2017Rotatable} of Hiroshima Synchrotron Radiation Center.  The crystals were cleaved in ultra-high vacuum and measured at a temperature of T=20 K. The band structure calculation is performed using the VASP code \cite{Kresse_vasp_1993} and the Fermi surface is calculated by Wannier interpolation using wannier90 \cite{mostofi_wannier90_2014}. The phonon dispersion is calculated by quantum espresso \cite{giannozzi_quantum_2009} with norm-conserving potential on a q-grid $10\times6\times4$.

CuTe has an orthorhombic structure with space group $\it Pmmn$ and lattice constants of $\it a$ = 3.149 $\AA$, $\it b$ = 4.086 $\AA$, $\it c$ = 6.946 $\AA$  \cite{CuTeJACS1994}. The X-ray and Laue diffraction patterns of the CuTe sample at room temperature in the ab plane are shown in Fig.~S1(a, b) in the Supplementary Materials \cite{Supplemental}. For the non-CDW structure, evenly-spaced Te chains run above and below the puckered copper layers, and each copper atom is coordinated with four Te atoms  (Fig.~\ref{Fig1}(a)). The crystal naturally cleaves between adjacent Te layers, which is ideal for low energy electron diffraction (LEED) and ARPES measurements. At low temperature, units of two or three Te atoms (marked by red broken lines) with shorter bond length (movement of atoms indicated by black arrows in Fig.~\ref{Fig1}(a)) alternate with single Te atom along the $\it a$ axis with concomitant slight distortion of corrugated Cu layers along the $\it c$ axis direction. This results in a 5$\times$1$\times$2 superstructure \cite{Stolze2013CuTe} with projected modulation wave vectors of $q_x$ = 0.4 a$^*$ and $q_z$ = 0.5 c$^*$ along the x and z directions respectively. The CDW is revealed by the 5$\times$1 reconstruction peaks (red circles in Fig.~\ref{Fig1}(b)) in the LEED pattern and a dip in transport measurement (Fig.~\ref{Fig1}(c)).  Figure \ref{Fig1}(d) shows the bulk Brillouin zone (BZ) with the projected x and z components of the modulation wave vector. The calculated Fermi surface for the normal phase of CuTe is shown in Fig.~\ref{Fig1}(d) with high symmetry points labeled.

\begin{figure*}
\centering
\includegraphics[width=17.8 cm] {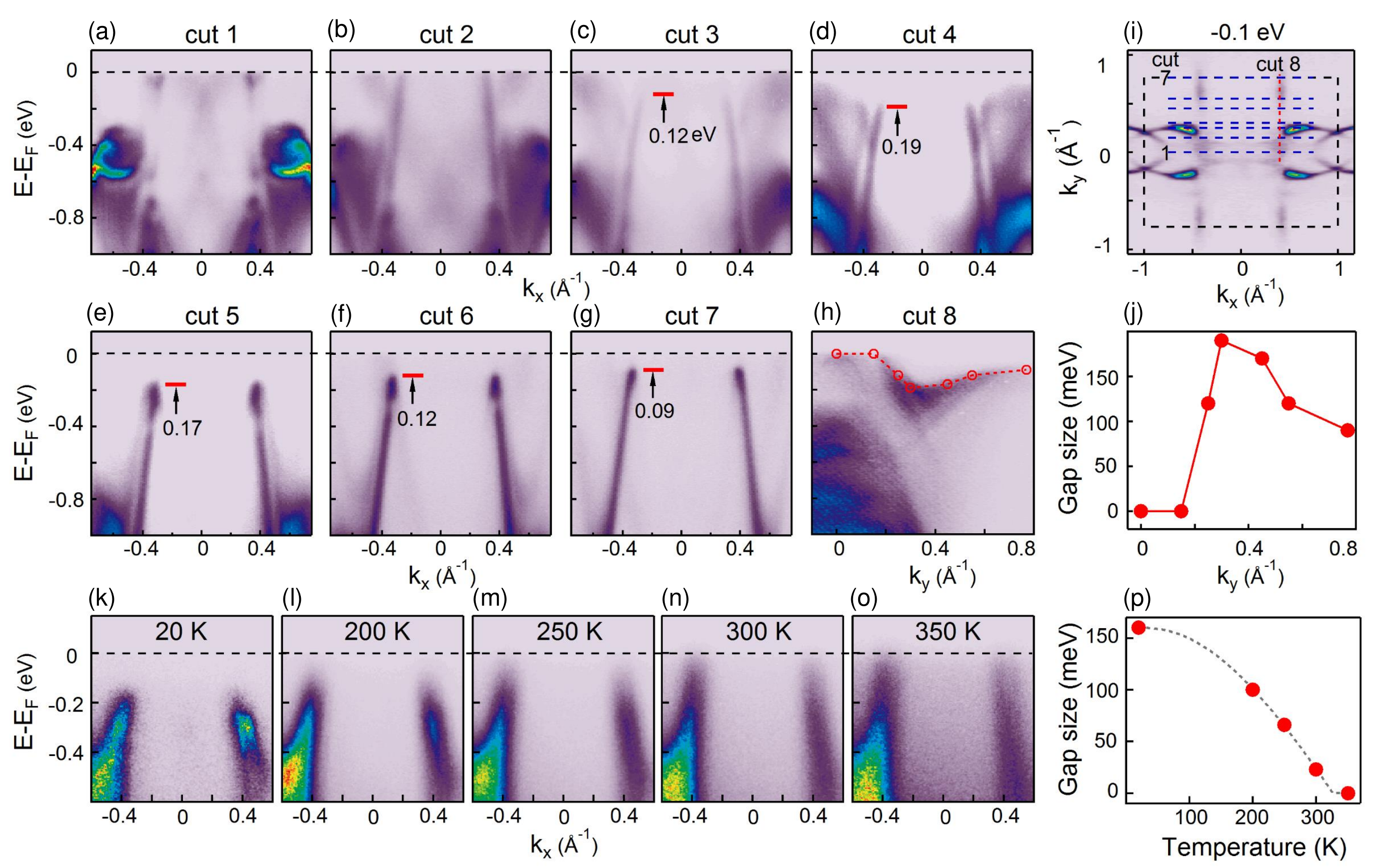}
\label{Fig2}
\caption{ \textbf{(a-g)} Band dispersions along the k$_x$ direction at fixed k$_y$ = 0, 0.15, 0.25, 0.3, 0.45, 0.55, 0.77 $\AA^{-1}$, respectively. The CDW gap size is labeled. \textbf{(h)} Band dispersion along the quasi-1D line segment at k$_x$ = 0.4 $\AA^{-1}$ .  The locations of (a-h) momentum cuts are marked by blue dashed lines in (i). Gap size extracted from (a-g) are appended as symbols. \textbf{(j)} Extracted gap size as a function of k$_y$. The error bars are smaller than the markers. \textbf{(k-o)} Evolution of band dispersions at k$_y$ = 0.47 $\AA^{-1}$ with temperature from 20 K to 350 K with photon energy of 60 eV. \textbf{(p)} Extracted gap size as a function of temperature. The gray dashed line is the fitting curve using a BCS-type gap equation.}\label{Fig2}
\end{figure*}

Figure 1(e-j) shows ARPES intensity maps measured at 20 K from the Fermi energy (E$_F$) to -0.5 eV with p polarized light. Intensity maps with the s polarized light are shown in Fig.~S3 in the Supplementary Materials \cite{Supplemental}. The Fermi surface map shows pockets extending along the k$_x$ direction which are contributed by the $p_y$ orbitals (yellow, green and red pockets in Fig.~\ref{Fig1}(d), or the calculated intensity maps in the non-CDW state in Fig.~S2 of the Supplemental Material \cite{Supplemental} for more details), while the quasi-1D line segments along the k$_y$ direction (labeled as $\alpha$ and $\beta$ in Fig.~\ref{Fig1}(d)), which are contributed by the $p_x$ orbital of Te, are totally missing in the Fermi surface map, suggesting that they are gapped. At -0.1 eV, part of the quasi-1D line segments becomes observable and the gapped region becomes smaller at lower energy (see red arrows for the edges of the line segment). This indicates that the gap closes in some momentum regions and suggests that the gap size is rather anisotropic.  At -0.2 eV, the gap is fully closed and the entire line segments become detectable.  The separation between these line segments is q$_N$ $\approx$  0.4 a$^*$, which matches well with the projected in-plane component of the CDW wave vector  $q_x$ \cite{Stolze2013CuTe}. Therefore, the gapped quasi-1D line segments from the Te $p_x$ orbital and the matching CDW vector provide a direct experimental evidence for the CDW formation at low temperature.

 \begin{figure*}
\centering
\includegraphics[width=17.8 cm] {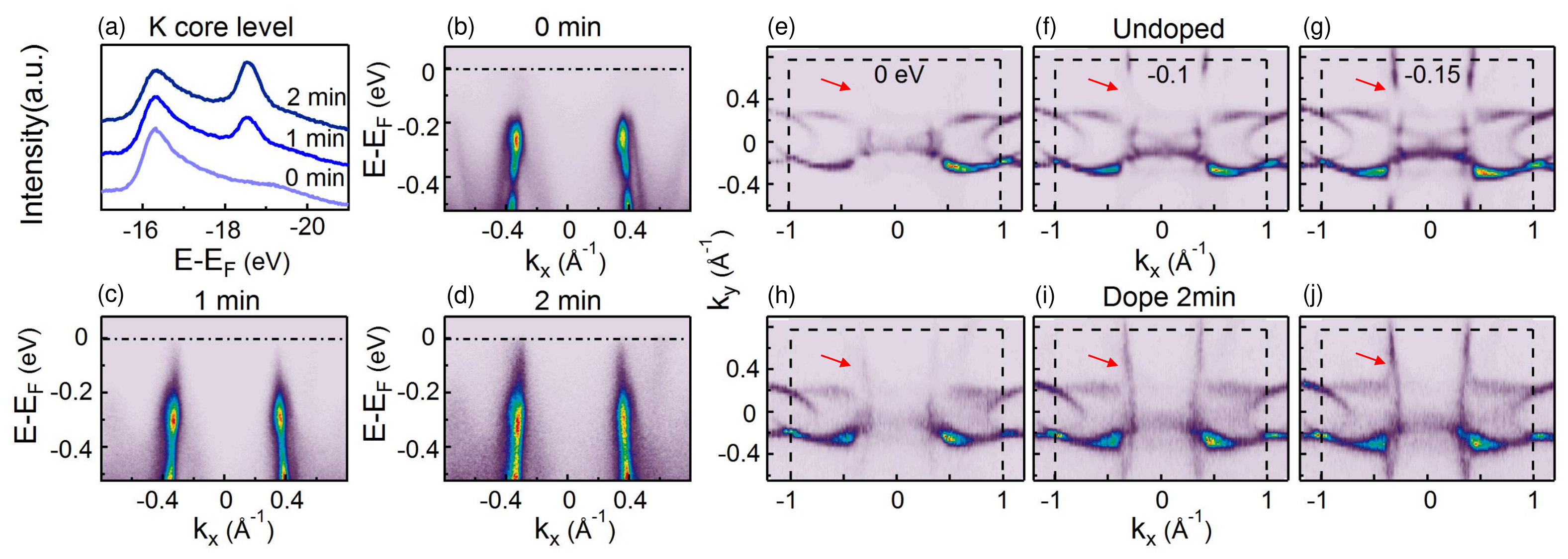}
\label{Fig3}
\caption{\textbf{(a)} Evolution of K 3p core-level spectra upon surface deposition of K atoms. \textbf{(b-d)} Band structure evolution of CuTe at k$_y$ = 0.45 $\AA^{-1}$ (cut 5 in Fig.~2(i)) with potassium doping at different time. \textbf{(e-g)} Constant energy maps from E$_F$ to -0.15 eV before doping at 20 K. \textbf{(h-j)} Constant energy maps from E$_F$ to -0.15 eV after doping at 20 K. The constant energy maps in Fig.~3 were measured by rotating the sample plane (azimuthal angle) by 90 degrees relative to Fig.~1 in order to align the analyser slit parallel to cut 5 direction, while maintaining other experimental conditions.}\label{Fig3}
\end{figure*}

Since the gap is anisotropic along the quasi-1D line segment, we further investigate the evolution of the gap with momentum. Figure 2(a-g) shows the dispersions measured parallel to $k_x$ at different $k_y$ positions as marked in Fig.~2(i). Near k$_y$ = 0 (cuts 1 and 2), the dispersing bands are gapless and cross E$_F$.  Cut 3 starts to slice through the line segment, and a gap of 0.12 eV is observed. The gap increases and reaches the maximum value of 0.19 eV for cut 4 at k$_y$ = 0.3 $\AA^{-1}$, and eventually decreases to 0.09 eV at the BZ boundary for cut 7. The evolution of the gap size is in agreement with intensity maps shown in Fig.~1, where the line segment is only partially observed between E$_F$ and -0.15 eV and becomes completely visible at -0.2 eV when the gap fully closes. The gap anisotropy can also be directly visualized by slicing along the quasi-1D line segment at k$_x$ = 0.4 $\AA^{-1}$ (cut 8 shown in Fig.~2(h)). The dispersion along the line segment suggests that this band still has some extent of two-dimensional characteristics caused by the hybridization of the Te $p_x$ orbital with other orbitals.  This quasi-1D band is gapless around the $\Gamma$ point and reaches the maximum value at k$_y$ = 0.3 $\AA^{-1}$, corresponding to the bottom of the dispersing band in Fig.~2(h). Moving toward the BZ boundary, the gap size decreases gradually, corresponding to the slow rise of the dispersion towards E$_F$ in (h). The momentum dependence of the gap size is in overall agreement with the calculated band dispersion with a $5\times1\times2$ supercell in Fig.~S4 in the Supplementary Materials \cite{Supplemental}. Such momentum dependence is attributed to the imperfect nesting, and is in consistent with the nesting-driven scenario as observed in other CDW system \cite{Lou2016Interplay}. The folded band can also be observed in the dispersion by adjusting the color contrast of the data shown in Fig.~2(f) (see Fig.~S5 in the Supplementary Materials \cite{Supplemental}). Taking the maximum gap value of 0.19 eV and T$_c$ = 335 K, we obtain the ratio of $2\Delta/k_B T_c$ $\approx$ 13, which is similar to that reported in other CDW materials, e.g. 2H-TaSe$_2$ \cite{Rossnagel2005Fermi}, and is much larger than the 3.52 expected from mean-field theory for a weakly coupled CDW system \cite{tinkham2004introduction}.

Figure 2(k-o) shows the temperature evolution of the dispersions at k$_y$ = 0.47 $\AA^{-1}$ (similar to cut 5), where the gap is close to maximum. With increasing temperature, the top of the valence band moves up toward E$_F$, indicating a decrease of the CDW gap size. The temperature dependent energy distribution curves (EDCs) and corresponding reference spectra are shown in Fig.~S6 in the Supplementary Materials \cite{Supplemental}, and the extracted gap size from the shift of the leading edge is plotted in Fig.~\ref{Fig2}(p). We have also included the normalized EDCs in Fig.~S7 in the Supplementary Materials \cite{Supplemental}, which can be used to directly visualize the evolution of spectral weight distribution with temperature. The normalized EDCs show that there is a spectral weight redistribution from lower binding energies near E$_F$ to higher binding energies, and the total spectral weight is conserved during the CDW transition. The extracted gap size in  Fig.~\ref{Fig2}(p) fits the semi-phenomenological BCS gap equation in the mean field theory $\bigtriangleup^{2}$(T) - $\bigtriangleup^{2}(T_c)$ $\propto$  tanh$^2$(A$\sqrt{T_c / T-1}$) \cite{Chen2015Charge}, where A is a fitting constant. The temperature dependence of the CDW gap confirms that the CDW transition is in agreement with the reported transition temperature from resistivity measurements, and suggests that quantum fluctuation effects are negligible compared with thermal fluctuation effects considering the high T$_c$ value for this system.

\begin{figure}
\flushleft
\includegraphics[width=8.6 cm] {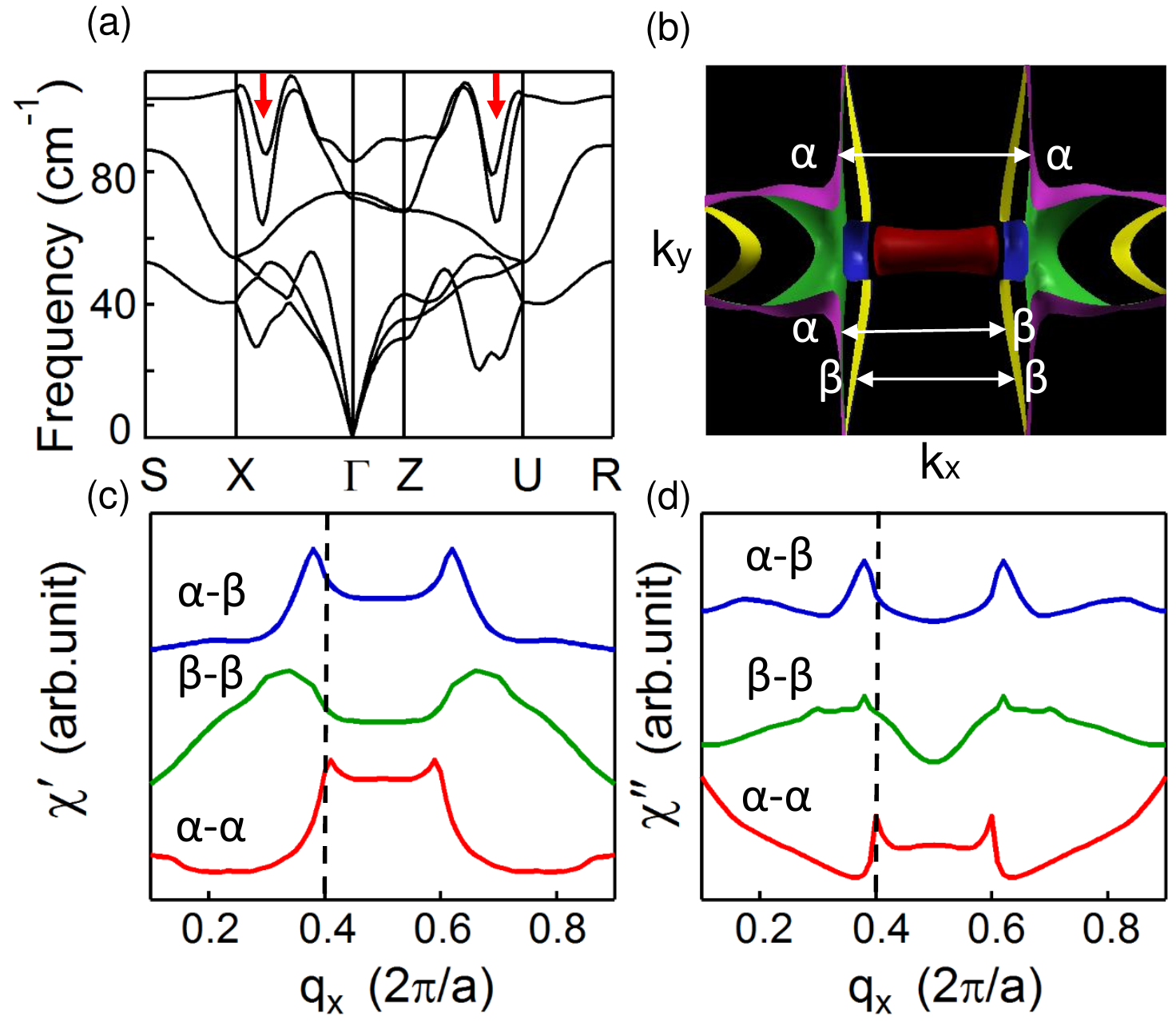}
\label{Fig4}
\caption{\textbf{(a)} Calculated phonon spectrum along the high symmetry directions. The red arrows indicate the Kohn Anomaly along the $\Gamma$-X and Z-U directions. \textbf{(b)} Schematic drawing of scattering between $\alpha$ and $\beta$ bands. \textbf{(c, d)} Real-part and Imaginary-part of the susceptibility between different bands along the $\Gamma$-X direction.}\label{Fig4}
\end{figure}

In addition to increasing tempeature, the CDW gap can in principle also be filled by injection of charge carries. For examples, transient melting of the CDW phase has been reported in TbTe$_3$ \cite{ShenTbTe3TrARPES} and TiSe$_2$ \cite{TiSe2Nature,TiSe2NatCommun} through photoexcitation by ultrashort laser pulses. Here we investigate the collapse of the CDW gap upon electron doping by {\it in situ} surface deposition of potassium (K) on CuTe.  The successful deposition is confirmed by the K 3p core-level peaks at -18.5 eV as shown in Fig.~\ref{Fig3}(a). Figure \ref{Fig3}(b-d) shows the band structure evolution of CuTe at k$_y$ = 0.45 $\AA^{-1}$ as we deposit K on the sample surface. The gapped electronic state is gradually filled by doped electrons and eventually the band reaches E$_F$ (Fig.~\ref{Fig3}(d)), indicating the closing of the CDW gap. A comparison of constant energy maps for undoped (Fig.~\ref{Fig3}(e-g)) and doped (Fig.~\ref{Fig3}(h-j)) surfaces show that the originally gapped region in the undoped sample becomes gapless after doping (marked by arrows), indicating that injection of electrons by surface K doping destroys the CDW gapped state.

First principles calculations of the phonon dispersion and the electron susceptibility have been performed to provide theoretical insights to the CDW mechanism.  A phonon softening is observed at both $q=(0.4,0,0)$ and $(0.4,0,0.5)$ (pointed by red arrows) in the calculated phonon spectra shown in Fig.~\ref{Fig4}(a), suggesting that it does not strongly depend on k$_z$. Such Kohn anomaly coincides with the $q_{CDW}$, suggesting electron-phonon coupling as a possible origin for the CDW. Another possible origin for CDW formation, namely Fermi surface nesting, can be evaluated by the calculated real ($\chi^\prime$, diverging for triggering CDW formation) and imaginary ($\chi''$, reflecting the Fermi surface topology) parts of the electron scattering susceptibility \cite{M2008Fermi} according to $\chi'=\sum_k\frac{f(\epsilon_k)-f{\epsilon_{k+q}}}{\epsilon_k-\epsilon_{k+q}}$ and $\lim_{\omega->0}\frac{\chi''(\omega)}{\omega}=\sum_k \delta(\epsilon_k-\epsilon_F)\delta(\epsilon_{k+q}-\epsilon_F)$, where $f(\epsilon_k)$ is the occupation at energy $\epsilon_k$.
For both of two quasi-1D bands from the Te $p_x$ orbitals (labeled as $\alpha$ and $\beta$ in Fig.~4(b)), a peak is observed at $q_x\approx 0.4$ a$^*$ and the best match for $q_{CDW}$ = 0.4 a$^*$ is obtained for scattering between the straight $\alpha$ bands (Fig.~4(c, d)), indicating that the contribution of nesting is mainly derived from $\alpha$-$\alpha$ scattering. Such Fermi surface nesting with high electron susceptibility can favour CDW formation at $q_x$ = 0.4 $a^*$.  Therefore, both electron-phonon interaction and electron-electron scattering due to Fermi surface nesting are possible origins of the CDW phase formation in the ground state of CuTe.

In summary, we report the electronic structure and provide experimental evidence for the CDW phase in CuTe containing quasi-1D Te chains. ARPES experiments clearly show that the two quasi-1D like parallel bands from the Te $p_x$ orbitals are gapped in CDW transition, and the gap is collapsed by increasing temperature or electron doping. Our experimental results together with first principles calculations suggest that electron-phonon coupling and electron-electron scattering by Fermi surface nesting of the quasi-1D Te $p_x$ orbitals  both contribute to the formation of the CDW phase of CuTe.

This work is supported by the National Natural Science Foundation of China (Grant No.~ 11725418 and 11334006), Ministry of Science and Technology of China (Grant No.~2016YFA0301004, 2016YFA0301001 and 2015CB921001), Science Challenge Project ( No. TZ2016004) and Beijing Advanced Innovation Center for Future Chip (ICFC). This research used resources of the Advanced Light Source, which is a DOE Office of Science User Facility under contract No. DE-AC02-05CH11231. Experiments at HiSOR were performed through Proposals No.~16BU001.

\bibliography{reference}

\end{document}